\documentclass[12pt]{revtex4-1}
\usepackage{graphics,graphicx,epsfig,latexsym,amssymb,verbatim,color,bm}
\usepackage[caption=false]{subfig}
\usepackage{graphicx}
\usepackage{amsfonts}
\usepackage{amsmath}
\usepackage{amsbsy}
\usepackage{array}
\usepackage{dsfont}

\newcommand{\be}{\begin{eqnarray}}
\newcommand{\ee}{\end{eqnarray}}

\usepackage{theorem}

\def\squareforqed{\hbox{\rlap{$\sqcap$}$\sqcup$}}
\def\qed{\ifmmode\squareforqed\else{\unskip\nobreak\hfil
\penalty50\hskip1em\null\nobreak\hfil\squareforqed
\parfillskip=0pt\finalhyphendemerits=0\endgraf}\fi}
\def\endenv{\ifmmode\;\else{\unskip\nobreak\hfil
\penalty50\hskip1em\null\nobreak\hfil\;
\parfillskip=0pt\finalhyphendemerits=0\endgraf}\fi}

\begin{document}

\title{Geometrical Bell inequalities for arbitrarily many qudits with different outcome strategies}

\author{Marcin Wie\'{s}niak, Arijit Dutta, and Junghee Ryu}
\address{Institute of Theoretical Physics and Astrophysics, University of Gda\'{n}sk, 80-952 Gda\'{n}sk, Poland}

\begin{abstract}
Greenberger-Horne-Zeilinger (GHZ) states are intuitively known to be the most nonclassical ones. They lead to the most radically nonclassical behavior of three or more entangled quantum subsystems. In case of two-dimensional systems, it has been shown that GHZ states lead to an exponentially higher robustness of Bell nonclassicality against the white noise in case of geometrical inequalities than in case of WWW\.ZB ones. We introduce geometrical Bell inequalities (BIs) for collections of arbitrarily many systems of any dimensionality. We show that the violation factor of these inequalities grows exponentially with the number of parties and study their behavior in function of dimensionality of subsystems and number of local measurements. We also investigate various strategies of assigning mathematical objects to events in the experiment, each leading different violation ratios.
\end{abstract}



\maketitle
\newcommand{\bra}[1]{\left\langle#1\right\vert}
\newcommand{\ket}[1]{\left\vert#1\right\rangle}
\newcommand{\abs}[1]{\left\vert#1\right\vert}
\newcommand{\avg}[1]{\langle#1\rangle}
\newcommand{\pro}[2]{|\left<#1|#2\right>|^2}
\newcommand{\tmp}[2]{\left\vert\langle#1\right\vert#2\left\rangle\vert^2\right}
\newcommand{\hket}[1]{\left\vert#1\right)}
\newcommand{\textleft}[2]{{\vphantom{#2}}{_{#1}}#2}

\section{Introduction}
Potency of various states to violate Bell inequalities (BIs) \cite{BELL}, apart from its fundamental consequences,  distinguishes them as forms of a resource directly usable in quantum information processing. Not only can the violation ensure us about the security of a scheme of a cryptographic key generation \cite{Ekert}, but also it can provide communication advantage in distributed computing \cite{Zukowski1}, or increase security of secret sharing protocols \cite{Zukowski2}. It is hence an interesting and important question to investigate, in which situations such a violation can occur.

Various schemes of generating BIs for collections of qubits have been found (e.g., \cite{WW,WZ,ZB,WNZ,LMPW}). For larger subsystems, still very little is known about falsifying local hidden variable models in general. The most profound versions of the theorem are CGMLP inequalities \cite{CGLMP} and their chained generalizations  \cite{CHAINEDD}. From the experience with qubits we know that the violation ratio of BIs can grow with the number of parties involved in the experiment, thus we expect it to be the same for qudits. We would also like to believe that the higher dimensionality can lead to 
stronger non-classical effects. Similarly, one may also check if the contrast between quantum mechanics and local realistic models is more radical under a closer inspection of the system, i.e., with more measurement settings to choose from by each observer. Results for qubits lead to various conclusions \cite{Lask2}.

In this work, we present a new Bell scheme, in which these questions can be at least partially answered. Specifically, we formulate geometrical BIs for any number of subsystems $N\geq 2$, any number of local measurement settings $L\geq 2$, and any dimensionality of each local subsystem $d\geq 2$. Geometrical BIs have been introduced in Ref. \cite{planar}. They are based on an approach, in which the correlation function is a vector with components described by measurement settings. A scalar product of the quantum correlation function with itself is compared with the product with all local realistic models. In the original version, they have utilized all possible measurements lying in a given plane. This approach led to an observation that among all correlation-based inequalities, they provide the strongest robustness of the violation against the white noise. Subsequently, they were generalized to the case of finite number of measurements. 

Note that geometrical Bell inequalities were already formulated for qutrits in Ref. \cite{DUTTA}. The estimates have shown that, indeed, the large number of parties provides a stronger violation. These inequalities, however, do not fit our scheme.

The elasticity of the model give us also an opportunity to check what violation ratios are observed under various treatments of outcomes. We will compare three different strategies. In the first we will treat each local measurement as a dichotomic one. It will be said to yield one value if a single specific local outcome have occurred. Second, we will consider the case in which outcomes of local measurements are assigned integers, which are sent to a referee. The referee sums the results modulo the dimension of each subsystem. The third  treatment is to associate local outcomes with complex root of unity and compute the correlation function by their multiplication. We are thus able to see the performance of considering each probability, a specific type of correlations, or a commonly used straight-forward generalization of Pauli matrices as unitary operators based approach in the same Bell scheme (e.g., \cite{RYU13,RYU14}). This shall be seen as a hint for future constructions of optimal BIs. In this way we want to emphasize various degrees of ignorance introduced in constructing the correlation function.

\section{Geometric Approach}
Consider a real vector $\vec{V}$ and a set of real vectors $S$. Any element of $S$ will be denoted as $\vec{S}$. If the norm of $\vec{V}$ exceeds the scalar product with all elements of $S$, then it cannot belong to this set, i.e., $\vec{V}$ cannot be represented as a convex combination of elements of $S$:
\begin{equation}
\vec{V}\cdot\vec{V}>\vec{S}\cdot\vec{V}\Rightarrow \vec{V}\notin S.
\label{eq:geo_eq}
\end{equation}
Note, however, that the converse statement is not true in general.

As the vector $\vec{V}$ we take the quantum correlation function in the form of $E_{\mathrm{QM}}(\alpha_1, \alpha_2, \dots, \alpha_N)$, where $\alpha_j$ denotes a parameter of measurement observables for $j$th party. This function is the average of the product of local results. On the other hand, the local realistic (LR) theories assume that the local results are predetermined before the measurements, contrary to the quantum mechanical description. Then, the correlation function can be simulated by $E_{\mathrm{LR}}(\alpha_1, \alpha_2, \dots, \alpha_N)=\int d\lambda \rho(\lambda) I_{1}(\alpha_1,\lambda) I_{2}(\alpha_2,\lambda) \cdots I_{N}(\alpha_N,\lambda)$, where $\lambda$ represent hidden variables, $\rho(\lambda)$ is a probability distribution, and $I_{j}(\alpha_j,\lambda)$ is the predetermined results of the measurement observables. The $\vec{S}$ will be the LR correlation function. In other words, if the correlation functions $E_{\mathrm{QM}}$ and $E_{\mathrm{LR}}$ holds in (\ref{eq:geo_eq}) for all $E_{\mathrm{LR}}$, then the LR description cannot describe the quantum prediction. With the different outcome strategies, we numerically calculate the ratio between quantum and classical description in the form of $(E_{\mathrm{QM}}\cdot E_{\mathrm{QM}})/\mathrm{max}_{E_{\mathrm{LR}}}(E_{\mathrm{QM}}\cdot E_{\mathrm{LR}})$, we call it a quantum-to-classical ratio (QCR).


Here, we will exploit the principle known for qutrits as the 1-0-1 rule~\cite{DUTTA}. It states that for three squared orthogonal components of spin-1 the outcomes of measurements will be $1,1$, and $0$ in some order. However, in noncontextual and local theories, the assignment of $0$ to a specific state cannot change if a compatible measurement is performed. It hence lies at the heart of Kochen-Specker and Bell arguments. We are interested in such orbits of observables, which involve commuting operators, but with changed (permuted) eigenvalues. 

Given the dimensionality of each subsystem $d$, the permutation of these vectors is easily realized by the following transformation
\begin{eqnarray}
U(\alpha)=\mathrm{diag}(1, e^{i \alpha}, e^{2 i \alpha}, \dots, e^{(d-1) i \alpha}) \, F,
\end{eqnarray}
where $\mathrm{diag}(\cdot)$ is a diagonal matrix and $F$ is the Fourier transform,
\begin{equation}
F=\frac{1}{\sqrt{d}}\left(
\begin{array}{cccc}
1&1&1& \\1&\omega&\omega^2&\cdots \\1&\omega^2&\omega^4&\\ &\vdots&&\ddots
\end{array}
\right),
\end{equation}
and $\omega=\exp(2\pi i/d)$. Notice that $U(\alpha)$ realizes cyclic permutations for $\alpha=2k\pi/d$ with integers $k$.

\section{Outocme Strategies}
\subsection{Strategy I : Multiplying real local outcomes}

We first consider the case of traditional von Neumann measurements, where the outcomes are simply scalars and the correlation function is the expectation value of the local outcomes. We study these observables, for which one-dimensional subspace is distinguished from the rest by an eigenvalue:
\begin{equation}
J(\alpha)=U(\alpha)\mathrm{diag} \left( \frac{d-1}{d},\frac{-1}{d},\dots,\frac{-1}{d} \right) U^\dagger(\alpha).
\end{equation}
Let us remark that all strategies will involve outcomes, which sum up to $0$. Only then can we associate the ratio between the quantum and the maximum LR value with a strength of violation and robustness against the white noise. This is due to the fact that since our local observables are traceless. Then, for the generalized Werner states in the form of $\rho=p\ket{\psi}\bra{\psi}+(1-p)\openone/d^N$, a convex combination of an entangled pure state $\ket{\psi}$ and white noise, the mean values are scaled by factor $p$ (no contribution from the white noise due to the traceless). This results in the left-hand side of the geometrical condition (\ref{eq:geo_eq}) scales as $p^2$, while the right-hand side as $p$.

The state under the consideration is the generalized Greenberger-Horne-Zeilinger (GHZ) state of $N$ qu$d$its,
\begin{eqnarray}
\label{GHZ}
|\Psi_{d}^{N}\rangle=\frac{1}{\sqrt{d}}\sum_{j=0}^{d-1}\ket{j}^{\otimes N}.
\label{eq:generalized_ghz}
\end{eqnarray}
Then the quantum correlation function reads
\begin{eqnarray}
\label{Eqm}
E_{\mathrm{QM}}(\alpha_1,\dots,\alpha_N)=\langle \Psi_{d}^{N}|J(\alpha_1)\otimes \cdots \otimes J(\alpha_N)|\Psi_{d}^{N}\rangle.
\end{eqnarray}
Now, let us consider the fixed part of $J(\alpha)$ and $F \, \mathrm{diag}(\frac{d-1}{d},-\frac{1}{d},\dots,-\frac{1}{d}) F^\dagger$, which has the following matrix representation:
\begin{equation}
F|0\rangle\langle 0|F^\dagger-\frac{\openone}{d}=\frac{1}{d}\left(\begin{array}{cccc}0&1&1&\cdots \\1&0&1&\cdots \\1&1&0&\cdots \\ \vdots&\vdots&\vdots&\end{array}\right).
\end{equation} 
Because the GHZ state is correlated in the computational basis (all qu$d$its always yield the same outcome of a measurement in this basis), the correlation function can be written as
\begin{equation}
E_{\mathrm{QM}}(\alpha_1,\dots,\alpha_N)=\frac{2}{d^{N+1}}\sum_{j=1}^{d-1}(d-j)\cos (j\tilde{\alpha}),
\label{eq:eqm_s1}
\end{equation}
where $\tilde{\alpha}=\sum_{k=1}^{N} \alpha_{k}$. 

\begin{figure*}
\centering
\subfloat[For $N=2$.]
{
\includegraphics[width=8cm]{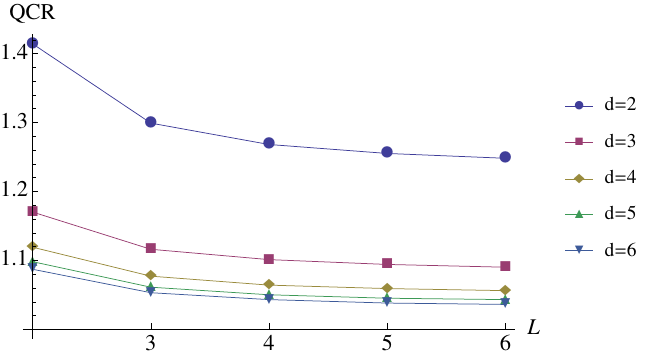}
\label{fig:N2QCR}
}\\
\subfloat[For $N=3$.]
{
\includegraphics[width=7cm]{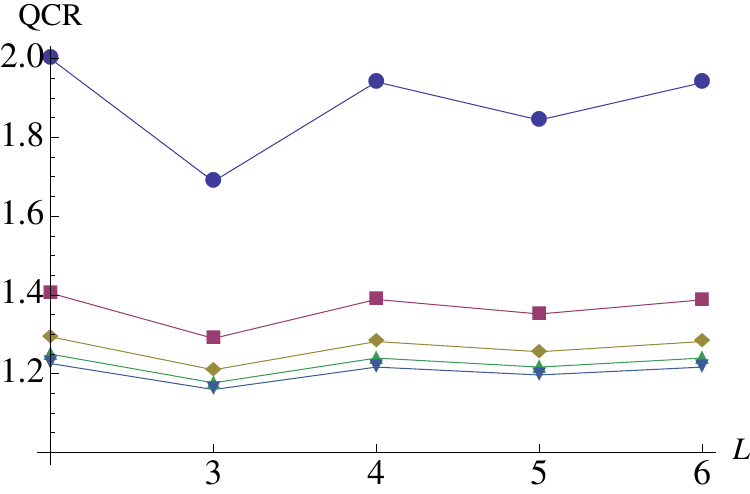}
\label{fig:N3QCR}
}
\subfloat[For $N=4$.]
{
\includegraphics[width=7cm]{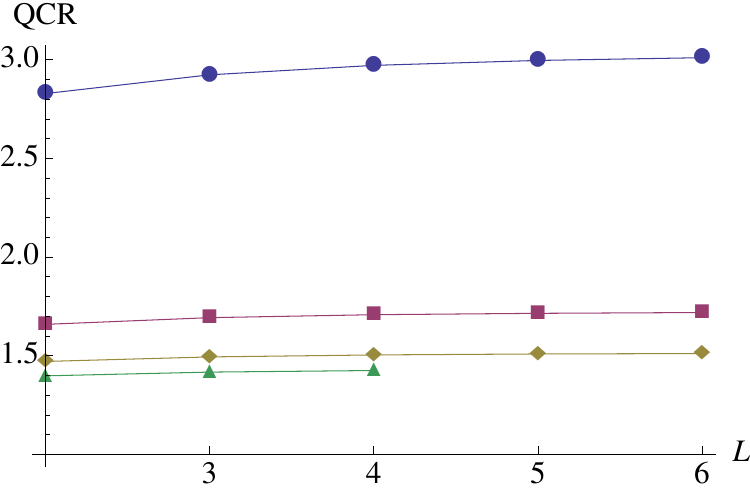}
\label{fig:N4QCR}
}
\caption{(Color online) The values of QCR (quantum to classical ratio) for the first and second strategies, assigning real values as outcomes, in the generalized GHZ state~(\ref{eq:generalized_ghz}) with $N=2, 3$ and $4$, respectively. They show the same results (See the main text for details). The numerical data are given in Table~\ref{tab:QCRreal} in \ref{appendix}.}
\end{figure*}

On the other hand, a LR model implies that the local outcomes are predetermined before measurements, i.e., the model freely preassigns values $(d-1)/d$ and $-1/d$. The only requirement is that once one setting $\alpha_i$ has been ascribed value $(d-1)/{d}$, we demand that all other settings $\alpha_j$ is assigned to $-{1}/{d}$. The settings used by each observer are $\alpha_i(j_i,k_i)=2\pi\left(\frac{k_i}{L}+j_i\right)/d$, with $L$ being the number of settings (i.e., different bases) for each observer. The variables $k_i \in \{0,1,...,L-1\}$ enumerates a basis $i$th observer measures in, and $j_i$ encodes the result the observer reports upon seeing one of his detectors triggered. 
Then the LR correlation function $E_{\mathrm{LR}}(\alpha_1,\dots, \alpha_N)$ is then simply a product of values assigned to specific settings. We numerically show the values of QCR for some combinations of $N,d$, and $L$ in Figs.~\ref{fig:N2QCR}, \ref{fig:N3QCR} and \ref{fig:N4QCR}, minimized over all LR models.

A few remarks ought to be made on these results. First is that, similarly to Ref. \cite{DUTTA}, we observe a sort of fluctuation of the values of QCR; the violation for $L=3$ is the lowest one and for for $L=5$ is the second (see Fig.~\ref{fig:N3QCR}). Let us mention that in Ref. \cite{DUTTA} another LR model turned out to be more optimal occasionally, whereas this is not the case here. Second, we observe the effect known from Ref. \cite{Lask1}; the violation ratio is the highest for two settings for two and three qudits (see Figs.~\ref{fig:N2QCR} and \ref{fig:N3QCR}), but for four (presumably more) parties the values grow with $L$, saturating at the limit of $L\rightarrow\infty$ (see Fig.~\ref{fig:N4QCR}). Finally, we should remark that the case of $d=L=2$ recovers the Mermin inequality \cite{Mermin}.      

We are also able to find the violation of the inequalities in the limit of $L\rightarrow\infty$. The quantum side of the inequality reads
\begin{eqnarray}
\label{eqmeqm}
E_{\mathrm{QM}}\cdot E_{\mathrm{QM}}
&=&\int_{0}^{2\pi} \cdots \int_{0}^{2\pi}E^2_{\mathrm{QM}}(\alpha_1,\dots,\alpha_N)d\alpha_1 \cdots d\alpha_N \nonumber \\
&=&(2\pi)^N\frac{(2d-1)(d-1)}{3d^{2N-1}}.
\end{eqnarray} 
On the other hand, since the $E_{\mathrm{QM}}$ in Eq. (\ref{eq:eqm_s1}) shows a global maximum at $\tilde{\alpha}=2\pi k$ with integer $k$, the optimal LR model $E_{\mathrm{LR}}$ can be obtained by integrating the correlation function over all $\alpha_i$ in intervals $(-\pi/{d},\pi/{d}]$. As a result, the QCR reads
\begin{eqnarray}
\label{continuous}
\mathrm{QCR}=\left(\frac{2\pi}{d}\right)^N\frac{(2d-1)(d-1)}{6\sum_{k=1}^{d-1}(d-k)\left[\frac{2 \sin (k\pi/d)}{k}\right]^N}.
\end{eqnarray}
For a derivation of this formula, see the \ref{appendixA}.

For $N=d=2$, this formula is in an agreement with the results of Ref. \cite{Nagata}. It also guaranties exponential growth of violation strength with the number of parties. For qubits, it behaves like $\left({\pi}/{2}\right)^N\approx 1.571^N$, for qutrits -- $\left({2\pi}/{(3\sqrt{3})}\right)^N\approx 1.209^N$, for ququats -- $\left({\pi}/{(2\sqrt{2})}\right)^N\approx 1.111^N$, and for qusexts ($d=6$) -- $\left({\pi}/{3}\right)^N\approx 1.047^N$. The violation growth factor tends to 1 as $d\rightarrow\infty$. Still, for any finite $d$ and $N$, we observe a firm violation.

\subsection{Strategy II : Summing local outcomes modulo $d$}

In the second scenario, clicking of each detector corresponds to an integer outcome, ordered in the increasing manner. After the measurements have been performed, the outcomes associated to the detectors that have clicked are sent to a referee, who sums them modulo $d$. If the sum is $0$ modulo $d$, then the value $(d-1)/d$ is taken as an outcome of the measurement. Otherwise, $-1/d$. Then, for the generalized GHZ state (\ref{eq:generalized_ghz}) the quantum mechanical correlation function $E_{\mathrm{QM}}$ is given by
\begin{equation}
E_{\mathrm{QM}}(\alpha_1,\alpha_2,\dots, \alpha_N)=\frac{2}{d^2}\sum_{j=1}^{d-1}(d-j)\cos\left(j \tilde{\alpha}\right),
\label{eq:q_corre_s2}
\end{equation}
where $\tilde{\alpha}=\sum_{k=1}^{N} \alpha_{k}$.

\begin{figure*}
\centering
\subfloat[For $N=2$.]
{
\includegraphics[width=8cm]{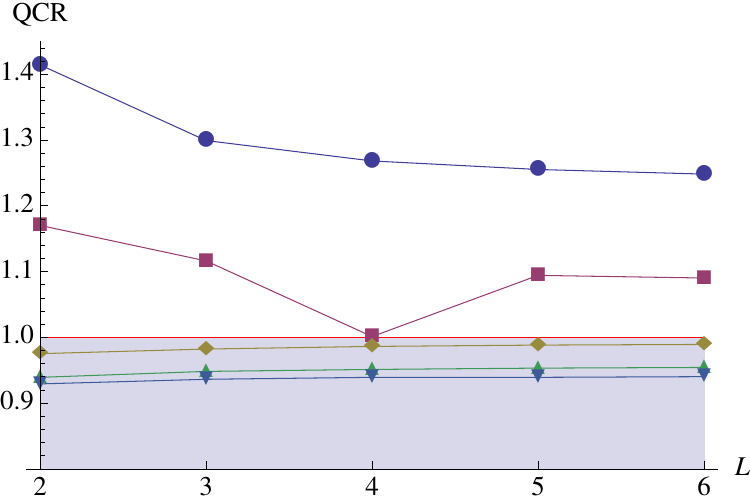}
\label{fig:N2C_QCR}
}
\subfloat[For $N=3$.]
{
\includegraphics[width=9cm]{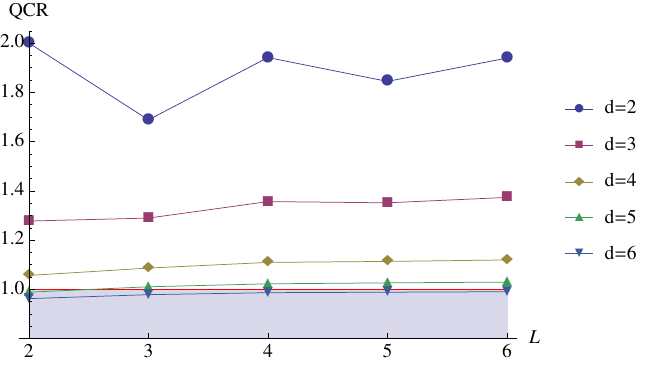}
\label{fig:N3C_QCR}
}
\caption{(Color online) The values of QCR (quantum to classical ratio) for the third strategy in the generalized GHZ state~(\ref{eq:generalized_ghz}) with $N=2$ and $3$, respectively. For the shaded region (QCR is less than one), no violations. The numerical data are given in Table~\ref{tab:QCRcomplex} in Appendix~\ref{appendix}.}
\label{fig:sample_subfloats}
\end{figure*}

This result is equivalent to the one~(\ref{eq:eqm_s1}) for the first strategy except for the coefficient. The LR correlation function is obtained similarly as for the first strategy. We show in Fig.~\ref{fig:sample_subfloats} that the values of QCR for the strategies I and II are equivalent for the generalized GHZ state for $N=2,3$, and $4$. This can be explained in the following. 

Notice that the span of the correlation function, which is the difference between its maximal and minimal values is 1. Also
\begin{equation}
\int_{0}^{2\pi}E_{\mathrm{QM}}(\alpha_1,\cdots,\alpha_N)d\alpha_i=0,
\end{equation} 
allows us to shift the local realistic model so that it takes only values 0 and 1. In such a case, a product with the optimal model would consist of an integral over $d^{N-1}$ boxes of dimension $\frac{2\pi}{d}\times \cdots \times\frac{2\pi}{d}$, representing sum of local outcomes $\left\{a_i:\sum_{i=1}^N a_i \,\, \mathrm{mod}\,\,d=0\right\}$. Each such box will centered at the peak of $E_{\mathrm{QM}}$, that is at $\tilde{\alpha}=2 \pi k$ with integer $k$.
\subsection{Strategy III: Multiplying complex local outcomes}
Another generalization of measurement outcomes on a qudit is $d$th order roots of unity over the complex field, $\omega=\exp(2\pi i /d)$. They combine the approach described above -- multiplying outcomes -- with those described below -- assigning objects to their sums modulo $d$. Again, upon an occurrence of a event (detector click) an observer reports outcome $0,...,d-1$. At each side, detectors are labeled in a natural manner. The referee compute $\omega^A$, where $A$ is the sum of the submitted outcomes. Then, the quantum correlation function for the GHZ state~(\ref{GHZ}) reads 
\begin{equation}
E_{\mathrm{QM}}(\alpha_1,\dots,\alpha_N)=\frac{1}{d}\left[(d-1)e^{-i\tilde{\alpha}}+e^{i(d-1)\tilde{\alpha}}\right],
\end{equation}
where $\tilde{\alpha} = \sum_{i=1}^{N} \alpha_i$.
In fact, we will compare $E_{\mathrm{QM}}\cdot E_{\mathrm{QM}}$ with $\mathrm{Re}(E_{\mathrm{LR}}\cdot E_{\mathrm{QM}})$. 
Note that the way that detector clicks are interpreted as outcomes implies a strong constrain on LR models. If a model assigns some outcome $a_i$ to angle $\alpha_i$, it must consequently assign $a_i-1$ to $\alpha_i+{2\pi}/{d}$, etc.. 

The procedure is the same as in other strategies; the inequalities for few low values of $d,L$, and $N$ are studied case-by-case to give us the idea about the structure of the optimal LR model. This turns out to assign a fixed outcome to an interval of angles, say, 0 to $-{\pi}/{d}\leq \alpha_i<{\pi}/{d}$, which we will call a packed model. Interestingly, for optimal  models, $(E_{\mathrm{LR}}\cdot E_{\mathrm{QM}})$ is strictly real. Violations for finite $L$ for $N=2,3$ are given in Tables \ref{tab:QCRreal} and \ref{tab:QCRcomplex}. In case of $L\rightarrow\infty$, we were able to find a formula for violation,
\begin{eqnarray}
\label{ViolComplex}
\mathrm{QCR}=\frac{(d-1)^2+1}{d\left[1+(d-1)^{N+1}\right]}\left[\frac{\pi (d-1)}{d \sin\frac{\pi}{d}}\right]^N.
\end{eqnarray}
Naturally, the violation is guaranteed for any $N$ for $d=2$, as the result from \cite{Nagata} must be recovered. However, for any finite $N$, the inequality is violated up to a certain value of $d$, and above that threshold, it is satisfied, tending to $1$ as $d\rightarrow\infty$. This is somewhat expected, since for large $d$, various powers of $\omega$ can be close to one another.  Numerical evaluation of Eq. (\ref{ViolComplex}) reveals that violation is sustained up to $d\approx 1.641N$

\section{Biased GHZ states}
\begin{figure}[h]
	\centering
		\includegraphics[width=9cm]{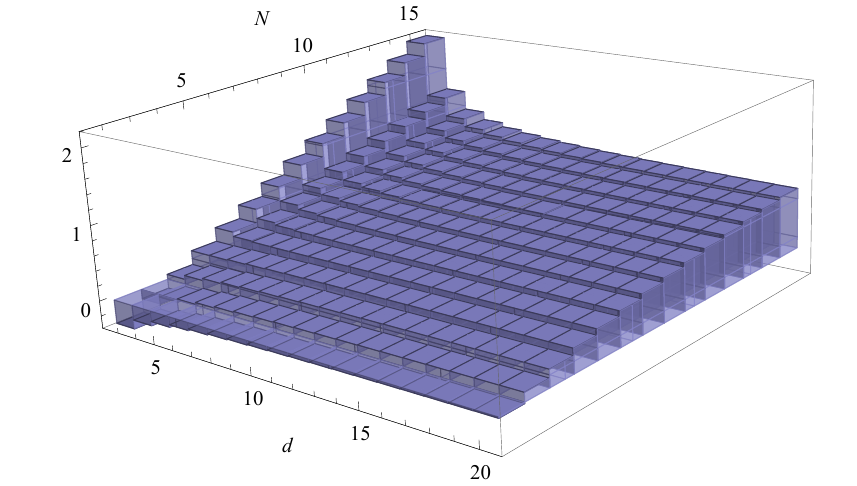}
	\caption{Logarithm of the right-hand side of Eq. (\ref{biased}) for $2\leq N\leq 15$  and $3\leq d\leq 20$.}
	\label{fig:Violation}
\end{figure}

Having found the first and the second strategies giving equivalent in the Bell scenario given above, we want to find a difference between them by altering the design of the experiment. Particularly, we consider biased GHZ states (of Schmidt rank $d-1$),
\begin{equation}	
|\psi\rangle=\frac{1}{\sqrt{d-1}}\sum_{i=1}^{d-1}|i\rangle^{\otimes N}.
\label{eq:bGHZ}
\end{equation}
There are few instances (e.g., $N=3,d=3$) where finite $L$ is optimal, but the inequality is not violated. First, let us consider the first strategy. In most cases, inequalities for $L\rightarrow\infty$ become optimal, and the violation ratio reads
\begin{eqnarray}
\mathrm{QCR}=\frac{(2d^2-7d+6)(2\pi)^N}{6d^N\sum_{k=1}^{d-1}(d-k-1)\left[\frac{2\sin{(k\pi/d)}}{k}\right]^N}.
\label{biased}
\end{eqnarray}
Note that for $N\leq 5$ the values of QCR increase with $d$, while for $N\geq 7$ they decrease. In case of $N=6$, the QCR is the lowest at $d=7$ (see FIG. \ref{fig:Violation}). 


The results distinguish the first strategy as the one leading to the highest robustness against the white noise. Once more, for low values of the particle number, $N<4$, it is optimal to refrain from measuring n more than two bases at each side. For $N\geq 4$ the QCR grows with $L$. Unlike for biased GHZ states, QCR grows with the dimensionality of the subsystems. The QCR for scalar dichotomic outcomes is given in Table \ref{Tab6} for $N=2,3$ and $L=2$.

When we assign complex outcomes, the correlation function becomes factorizable, $E_{\mathrm{QM}}=(d-2)/(d-1)e^{-i\alpha'}$ and hence the inequality cannot be violated.

\section{Conclusions}
We have presented geometrical Bell inequalities for a collection of qudits and an arbitrarily high number of local measurements settings. Their violation for the GHZ states has been demonstrated. Interestingly, regardless of $d$, in this state their violation is the strongest for $L=2$ distinct measurement bases with $N=2$ or 3 subsystems, and for $L\rightarrow\infty$ in case of $N\geq 4$. 

Within the same Bell scenario, we have also compared various strategies of treating the measurement outcomes. The fixed Bell experiment has guarantied us that the same amount of noncalssicality in the raw data. Basically, the aim to compare these strategies was to establish a degree of negligence we can afford to maintain the robustness against the white noise. 
First, we investigated reporting one value by an observer, when his/her specific detector clicks. Because of the symmetry of geometrical BIs, this revealed the full structure of the probability distribution. The other strategy was to sum up the outcomes of local measurements modulo $d$, while the last one was to represent this sum as one of complex roots of unity of degree $d$	. The last of these strategies represented measurement outcomes as numbers quite close to each other (for large $d$) and resulted in weak or no violation of BIs. 
The second singled out only a specific kind of correlation, and performed as good as the first one for the full rank GHZ states, but dropped back for biased ones.


Then we have decided to partially break the symmetry of the state by rejecting one of its Schmidt modes. We have been able to distinguish the real local scalars as the outcome strategy providing the most robust violation. This suggests that we can focus on general types of correlations, rather than individual probabilities only for highly symmetric states. Also, we shall point out that complex scalars have led to a fully factorizable correlation function. While it is possible to formulate all-versus-nothing paradoxes, in practical applications, complex measure outcomes can witness only the strongest correlations.




We would also like to stress that these BIs are relevant for analyzing the bright squeezed vacuum (BSV) state. The structure of each $n$-pair component of BSV is identical to the one of two-qudit singlets, and the unitary transformation can be conveniently realized with a polarization-dependent phase shift. However, an experimental challenge is to realize a projection on an unbiased superposition of all polarization states. Still, it might be possible to find similar inequalities utilizing projections more feasible in an experiment. In any case, BIs described here are yet another way to analyze BSV theoretically.

\section{Acknowledgements}
This work is a part of the project BRISQ2 financed by the European Commission. The work is subsidized form funds for science for years 2012-2015 
approved for
international co-financed project BRISQ2 by Polish Ministry of Science 
and Higher Education (MNiSW).
 The Authors acknowledge support from European Funds distributed by the Foundation for Polish Science (FNP). MW was initially supported within program HOMING PLUS, JR -- within TEAM, AD -- within MPD. MW was at the later stage supported by the Polish National Science Centre under grant DEC-2013/11/D/ST2/02638. JR was additionally supported by an ERC grant QOLAPS.

\appendix
\section{Derivation of formulae (\ref{continuous}) and (\ref{ViolComplex})}
\label{appendixA}
Let us begin with deriving the violation ratio for $L\rightarrow\infty$ for the first strategy, as given Eq. (\ref{continuous}). The numerator of this fraction has been already given in Eq. (\ref{eqmeqm}). Local observables have eigenvalues $(d-1)/d$ (unique) and $-1/d$ (degenerated). This means that any deterministic local model can be written as a product of following functions
\begin{equation}
I_{i}(\alpha_i)=\chi_i(\alpha_i)-\frac{1}{d},
\end{equation}
where $\chi_i(\alpha_i)=0,1$ is a characteristic function and, following from the 1-0-1 rule, $\int_0^{2\pi}\chi_i(\alpha_i)=2\pi/d$. When we calculate
\begin{eqnarray}
\max\limits_{E_{\mathrm{LR}}}(E_{\mathrm{QM}}\cdot E_{\mathrm{LR}})
&=&\max\limits_{\{I_{i}\}}\int_0^{2\pi}\int_0^{2\pi} d\alpha_1 d\alpha_2 \cdots \\ \nonumber 
&\times& E_{\mathrm{QM}}(\alpha_1,\alpha_2,...) I_1(\alpha_1)I_2(\alpha_2) \cdots,
\end{eqnarray}
we can neglect the constant part of each local realistic models. Thus we have
\begin{eqnarray}
\max\limits_{E_{\mathrm{LR}}}(E_{\mathrm{QM}}\cdot E_{\mathrm{LR}})
&=&\max\limits_{\{\chi_{i}\}}\int_0^{2\pi}\int_0^{2\pi} d\alpha_1 d\alpha_2 \cdots \\ \nonumber 
&\times& E_{\mathrm{QM}}(\alpha_1,\alpha_2,...) \chi_1(\alpha_1)\chi_2(\alpha_2) \cdots.
\end{eqnarray}
The $E_{\mathrm{QM}}$ has always a distinctive peak at $\tilde{\alpha}=0$, thus it is optimal to choose
\begin{equation}
\chi_i(\alpha_i)=\left\{\begin{array}{ll}1& -\frac{\pi}{d} \leq \alpha_i < \frac{\pi}{d}\\
0& \mbox{otherwise}\end{array}\right.
\end{equation}
which stright-forwardly leads to Eq. (\ref{continuous}). In the similar fashion, we find the same formula for strategy II. 

We therefore pass to Eq. (\ref{ViolComplex}). Numerical case-by-case studies show that the optimal local realistic model takes form
\begin{equation}
E_{\mathrm{LR}, \mathrm{opt}}(\alpha_i)=\left\{\begin{array}{ll}
1& -\frac{\pi}{d} \leq \tilde{\alpha} <\frac{\pi}{d}\\ 
\omega^{-1}   & \frac{\pi}{d} \leq \tilde{\alpha} <\frac{3\pi}{d}\\
\omega^{-2}   & \frac{3\pi}{d} \leq \tilde{\alpha} <\frac{5\pi}{d} \\
\vdots & \vdots \end{array}\right.
\end{equation}
Hence the integral can be divided into $d^N$ blocks of dimension $\frac{2\pi}{d}\times \cdots \times\frac{2\pi}{d}$. Each of them equally contributes to the integral, e.g.,
\begin{eqnarray}
\int_{-\frac{\pi}{d}}^{\frac{\pi}{d}} \int_{-\frac{\pi}{d}}^{\frac{\pi}{d}} &d\alpha_1& d\alpha_2 \cdots E_{\mathrm{QM}}(\alpha_1,...\alpha_N)E_{\mathrm{LR}, \mathrm{opt}}(\alpha_1,...\alpha_N) \nonumber\\
&=&\frac{2^N [1+(d-1)^N]\sin\left(\frac{2\pi}{d}\right)^N}{d(d-1)^N},
\end{eqnarray}
which straight-forwardly leads to Eq. (\ref{ViolComplex}).

\section{The values of QCR for geometrical BIs}
\label{appendix}

\begin{table}[p]
\subfloat[For $N=2$]{
\begin{tabular}{|c|c|c|c|c|c|}
		\hline
			&\multicolumn{5}{c|}{$L$}\\
			\hline
			$d$&2&3&4&5&6\\
			\hline
			2&1.414&1.299&1.268&1.255&1.248\\
			3&1.170&1.116&1.101&1.094&1.090\\
			4&1.119&1.077&1.064&1.059&1.056\\
			5&1.098&1.061&1.050&1.045&1.043\\
			6&1.087&1.053&1.043&1.038&1.036\\
			\hline
\end{tabular}
}
\qquad
\subfloat[For $N=3$]{
\begin{tabular}{|c|c|c|c|c|c|}
		\hline
			&\multicolumn{5}{c|}{$L$}\\
			\hline
			$d$&2&3&4&5&6\\
			\hline
			2&2&1.688&1.941&1.844&1.939\\
			3&1.404&1.289&1.388&1.351&1.387\\
			4&1.293&1.209&1.281&1.255&1.281\\
			5&1.249&1.176&1.239&1.216&1.239\\
			6&1.225&1.159&1.216&1.196&1.216\\
			\hline
		\end{tabular}
}		
\\
\subfloat[For $N=4$]{
\begin{tabular}{|c|c|c|c|c|c|}
		\hline
			&\multicolumn{5}{c|}{$L$}\\
			\hline
			$d$&2&3&4&5&6\\
			\hline
			2&2.828&2.923&2.971&2.996&3.010\\
			3&1.658&1.692&1.707&1.714&1.718\\
			4&1.470&1.493&1.503&1.508&1.510\\
			5&1.397&1.416&1.424&&\\
			\hline
		\end{tabular}
}
\caption{The values of QCR for the first and second strategies (assigning real outcomes) in the generalized GHZ state~(\ref{eq:generalized_ghz}) with $N=2,3$, and $4$, respectively. As previously stated in main text, they show the same results.}
\label{tab:QCRreal}
\end{table}

\begin{table*}[b]
	\subfloat[For $N=2$]{
		\begin{tabular}{|c|c|c|c|c|c|c|}
		\hline
			&\multicolumn{6}{c|}{$L$}\\
			\hline
			$d$&$2$&$3$&$4$&$5$&$6$&$\infty$\\
			\hline
			2&1.414&1.299&1.268&1.255&1.248&\\
			3&1.170&1.116&1.001&1.094&1.090&\\
			4&0.975&0.982&0.986&0.988&0.989&0.991\\
			5&0.939&0.948&0.951&0.953&0.954&0.956\\
			6&0.929&0.936&0.939&0.939&0.940&0.942\\
			\hline
		\end{tabular}
}
\subfloat[For $N=3$]{
		\begin{tabular}{|c|c|c|c|c|c|c|}
		\hline
			&\multicolumn{6}{c|}{$L$}\\
			\hline
			$d$&$2$&$3$&$4$&$5$&$6$&$\infty$\\
			\hline
			2&2.000&1.688&1.941&1.844&1.939&1.938\\
			3&1.277&1.289&1.356&1.351&1.373&1.387\\
			4&1.056&1.086&1.109&1.113&1.119&1.128\\
			5&0.988&1.010&1.022&1.026&1.029&1.034\\
			6&0.962&0.978&0.986&0.988&0.990&0.994\\
  		\hline
		\end{tabular}
		}
		\caption{The values of QCR for the third strategy (assigning complex values as outcomes) in the generalized GHZ state (5) with $N = 2$ and $3$, respectively.}
	\label{tab:QCRcomplex}
\end{table*}

\begin{table}[b]
		\begin{tabular}{|c|c|c|}
		\hline
		&\multicolumn{2}{c|}{$N$}\\
		\hline
		$d$&2&3\\
		\hline
		3&0.770&0.889\\
		4&0.863&0.976\\
		5&0.911&1.020\\
		6&0.940&1.047\\
		7&0.959&1.064\\
		8&0.973&1.077\\
		\hline
		\end{tabular}
	\caption{The values of QCR for scalar dichotomic outcomes in the biased GHZ states (\ref{eq:bGHZ}) with measurement setting $L=2$.}
	\label{Tab6}
\end{table}

\end{document}